\documentclass[aps,twocolumn]{revtex4}
\usepackage{graphicx}
\begin{document}
\preprint{}
\title{Warm asymmetric matter in the Quark Meson Coupling Model}
\author{P.K. Panda$^{1,2}$, G. Krein$^1$, D.P. Menezes$^2$ and 
C. Provid\^encia$^3$}
\affiliation{
$^1$Instituto de F\'{\i}sica Te\'orica, Universidade
Estadual Paulista, Rua Pamplona 145, 01405-900 S\~ao Paulo, SP, Brazil \\
$^2$Departmento de F\'{\i}sica, Universidade Federal de 
Santa Catarina, 88040-900 Florian\'opolis, SC, Brazil \\
$^3$Centro de F\'{\i}sica Te\'orica, Departmento de F\'{\i}sica, 
Universidade de Coimbra, Coimbra 3000, Portugal}
\begin{abstract}
In this work we study the warm equation of state of asymmetric nuclear matter 
in the quark meson coupling model which incorporates explicitly quark 
degrees of freedom, with quarks coupled to scalar, vector and isovector 
mesons. Mechanical and chemical instabilities are discussed as a function 
of density and isospin asymmetry. The binodal section, essential in the 
study of the liquid-gas phase transition is also constructed and discussed. 
The main results for the equation of state are compared with two common 
parametrizations used in the non-linear Walecka model and the differences are 
outlined.
\end{abstract}
\maketitle
\section{Introduction}
The understanding of the properties of hot dense nuclear matter is an important
problem in theoretical physics in the context of neutron stars~\cite{prakash} 
as well as in high-energy heavy-ion collisions experiments where nuclei undergo 
violent collisions~\cite{NA50}. Such collisions may produce states of nuclear 
matter at conditions far from those normally encountered in low energy 
collisions. For example, at high density or high temperature nucleons in 
the nucleus may dissolve into quarks and gluons. At low temperature, which 
can be attained in medium energy heavy-ion reactions, no such exotic state 
can be produced but there is the possibility of interesting liquid-gas 
phase transitions leading to the breakup of heated nuclei into small 
clusters or droplets of nucleons~\cite{multif}. Such phase transitions have 
been identified in multifragmentation experiments and in the crusts of neutron 
stars~\cite{pethick,pet93}. The possibilities of such phase transitions were 
previously considered by several authors using different 
approaches~\cite{barranco,muller,mp}. 

In the present paper, we employ the quark-meson coupling model 
(QMC)~\cite{guichon,ST} to investigate the liquid-gas phase transition in
nuclear matter at different isospin asymmetries. In the QMC model, nuclear 
matter is described as a system of non-overlapping MIT bags which interact 
through the effective scalar and vector mean fields, very much in the same 
way as in the Walecka model, or more generally Quantum Hadrodynamics 
(QHD)~\cite{qhd}. Many applications and extensions of the model have been 
made in the last years~\cite{ffactors,recent,temp,song}, including 
droplet formation at zero temperature~\cite{krein00}. An special improvement 
is related with a density dependent bag parameter. This model is known as 
modified QMC model (MQMC) \cite{jin96,mj98} and some of its characteristics are discussed in 
the last section of the present paper.

The crucial difference is that in the QMC, the effective meson fields 
couple directly to the confined quarks inside the nucleon bags instead to 
point like nucleons as in QHD. While the QMC model shares many similarities 
with QHD-type models, it however offers new opportunities for studying nuclear 
matter properties. One of the most attractive aspects of the model is that 
different phases of hadronic matter, from very low to very high baryon densities 
and temperatures, can be described within the same underlying model. In the
QMC, matter at low densities and temperatures is a system of nucleons 
interacting through meson fields, with quarks and gluons confined within 
MIT bags. For matter at 
very high density and/or temperature, one expects that baryons and 
mesons dissolve and the entire system of quarks and gluons becomes 
confined within a single, big MIT bag. Another important aspect of the 
QMC is that the internal structure of the nucleon is introduced explicitly. 
Although the internal structure of nucleons can be effectively taken 
into account with an effective field theory employing hadronic degrees 
of freedom in a derivative expansion~\cite{EFT},  it is clear that such 
an approach will fail to describe the transition to a quark-gluon phase. 
It is found that the equation of state (EOS) for infinite 
nuclear matter at zero temperature derived from the QMC model is much softer 
than the one obtained in  the Walecka model~\cite{qhd}. Also, the QMC nucleon 
effective mass lies in the range $ 0.7$ to $0.8$ of the free nucleon mass, 
which agrees with results derived from nonrelativistic analysis of scattering 
of neutrons from lead nuclei~\cite{mahaux87} and is  larger to the effective
nucleon mass in the Walecka model. At finite temperature, there arises yet 
another difference between predictions of QMC and QHD, namely the behavior 
of the effective nucleon mass 
with the temperature at fixed baryon density. While in QHD-type models the 
nucleon mass always decreases with temperature, in the QMC it increases. The 
difference arises because of the explicit treatment of the internal structure 
of the nucleon in the QMC. When the bag is heated up,  
quark-antiquark pairs are 
excited in the interior of the bag, increasing the internal energy of the bag.

Here, we analyze the QMC model at finite temperature for different proton 
fractions. The proper treatment of the problem with isospin asymmetry requires 
not only isoscalar scalar and vector mesons fields, but also isovector meson 
fields. In particular we investigate the consequences of the different thermal 
behavior of the nucleon mass in the QMC model on the liquid-gas phase 
transition. 
The results are also compared with the ones obtained with two commonly used 
parametrizations for a non-linear Walecka model (NLWM), namely, NL-3~\cite{nl3} 
and TM-1~\cite{tm1}. We organize the present paper as follows. In Section~II we 
describe the QMC model at finite temperature for asymmetric nuclear matter and 
discuss the binodal section. In Section~III we present our numerical 
results and discuss differences with other approaches. Our summary is presented 
in Section~IV.

\section{QMC at finite temperatures}
\subsection{Asymmetric matter at finite temperature}
Recently the QMC model has been generalized to finite temperature taking a
medium dependent bag constant~\cite{temp}. In this generalized model,
symmetric nuclear matter was considered and hence only the interaction of 
quarks through the exchange of effective isoscalar scalar ($\sigma$) and vector 
($\omega$) mesonic fields were taken into account. We now extend this model to 
asymmetric nuclear matter at finite temperatures and include the contribution 
of the isovector vector meson $\rho$, in addition to $\sigma$ and $\omega$.

In the QMC model, the nucleon in nuclear matter is assumed to be described by 
a static MIT bag in which quarks couple to effective meson fields, which are 
treated as classical in a mean field approximation. The quark field 
$\psi_q({\bf r},t)$ inside the bag then satisfies the Dirac equation
\begin{equation}
\Big[i{\vec \gamma}\cdot{\vec \partial}-(m_q^0-V_\sigma)-\gamma^0(V_\omega
+\frac{1}{2} \tau_{3q}V_\rho) \Big]\psi_q({\bf r},t)=0,
\end{equation}
where $V_\sigma=g_\sigma^q\sigma_0$, $V_\omega=g_\omega^q\omega_0$ and
$V_\rho=g_\rho^qb_{03}$
with $\sigma_0$, $\omega_0$ and $b_{03}$ being the classical meson
fields. $g_\sigma^q$, $g_\omega^q$ and $g_\rho^q$
are the quark couplings with the $\sigma$, $\omega$ and $\rho$ mesons
respectively. $m_q^0$ is the current quark mass and $\tau_{3q}$ is the
third component of the Pauli matrices. In the present paper we consider only 
nonstrange $q=u$ and $d$ quarks only.

At zero temperature, the normalized ground state (s-state) wave function 
for a quark in the bag is given as
\begin{equation}
\psi_q(\vec r,t)=N \exp(-i\frac{\epsilon_q t}{R})
\left(
\begin{array}{c} j_0(xr/R)
\\ i\beta_q \vec \sigma \cdot \hat r j_1(xr/R)
\end{array}
\right) \frac{\chi_q}{\sqrt {4\pi}} ,
\end{equation}
where
\begin{equation}
\beta_q=\sqrt{\frac{\Omega_q-Rm^*_q}{\Omega_q+Rm^*_q}},
\end{equation}
with
\begin{equation}
\Omega_q = \sqrt{x^ 2+(Rm^*_q)^ 2}, \hspace{1.0cm} 
m^*_q=m^0_q-g_\sigma^q\sigma ,
\end{equation}
and $\chi_q$ is a Pauli spinor.

At finite temperatures, the three quarks 
inside the bag can be thermally excited to higher states and also quark-antiquark 
pairs can be created. For simplicity, we assume that the bag describing the nucleon 
continues to remain in a spherical shape with radius $R$, which is now 
temperature dependent. The single-particle energies in units 
of $R^{-1}$ are given as
\begin{equation}
\epsilon_{q}^{n\kappa}=\Omega_q^{n\kappa}+ R(V_\omega \pm \frac{1}{2}V_\rho ),
\end{equation}
for the quarks and 
\begin{equation}
\epsilon_{\bar q}^{n\kappa}=\Omega_q^{n\kappa}- R(V_\omega \pm
\frac{1}{2}V_\rho ),
\end{equation}
for the antiquarks, where the $+$ sign is for $u$ quarks and $-$ for $d$ 
quarks, and 
\begin{equation}
\Omega^{n\kappa}_q = \sqrt{x^2_{n\kappa}+R^2{m^*_q}^2} .
\end{equation}
The eigenvalues $x_{n\kappa}$ for the state characterized by $n$ and $\kappa$
are determined by the boundary condition at the bag surface, 
\begin{equation}
i\gamma\cdot n \psi_q^{n\kappa}=\psi_q^{n\kappa} .
\end{equation}
Thus, the quark eigenvalues $x_{n\kappa}$ become modified by the surrounding 
nucleon medium at finite temperature.

The total energy from the quarks and antiquarks at finite temperature is 
\begin{equation}
E_{tot} = \sum_{q,n,\kappa}\frac{\Omega^{n\kappa}_q}{R}\left(f^q_{n\kappa} +
f^{\bar q}_{n\kappa}\right), 
\label{Etot}
\end{equation}
where
\begin{equation}
f^q_{n\kappa} =  \frac{1}{e^{(\Omega_{q}^{n\kappa}/R-\nu_q)/T}+1} ,
\label{fq}
\end{equation}
and
\begin{equation}
f^{\bar q}_{n\kappa} = \frac{1}{e^{(\Omega_{q}^{n\kappa}/R+\nu_q)/T}+1},
\label{faq}
\end{equation}
with $\nu_q$ being the effective quark chemical potential, related to the
true quark chemical potential $\mu_q$ as
\begin{equation}
\nu_q=\mu_q-V_\omega-m_\tau^q \, V_\rho .
\label{nuq}
\end{equation}
The bag energy now becomes
\begin{equation}
E_{bag}=E_{tot}-\frac{Z}{R}+\frac{4\pi}{3}R^3B,
\label{Ebag}
\end{equation}
where $B$ is the bag constant and $Z$ parametrizes the sum of the center-of-mass
(c.m.) motion and gluonic corrections. Note that this center of mass correction
is different from that of~\cite{jin96}.
The effective nucleon mass is obtained from the bag energy and reads:
\begin{equation}
M^*_N=E_{bag}.
\label{MN}
\end{equation}
The bag radius $R$ is determined by the equilibrium condition for the nucleon bag
in the medium
\begin{eqnarray}
\frac{\partial M^*_N}{\partial R}=0.
\label{MNR}
\end{eqnarray}
Once the bag radius is obtained, the effective nucleon mass is immediately 
determined. For a given temperature $T$ and scalar field $\sigma$, the 
effective quark chemical potentials $\nu_q,\, q=u,\, d$ are determined
from the total number of quarks of each type in the proton and in the neutron, 
i.e.,
\begin{eqnarray}
n^j_0 &=& \sum_{q,n,\kappa}\left(f^q_{nq} - f^{\bar q}_{nq}\right) 
\equiv 3, 
\label{muqa} \\
n^j_3 &=& \sum_{q,n,\kappa} \, 2m_{\tau(q)}\left(f^q_{nq} - 
f^{\bar q}_{nq}\right) \equiv 2 m_{\tau(j)},
\label{muqa1}
\end{eqnarray}
for $j=p,n$. 

From the last two expressions one can see that the masses 
for protons and neutrons are different, since the effective chemical 
potentials $\nu_q,\, q=u,d$ are not the same. Nevertheless, it is a 
reasonable approximation, for the purposes of the present paper, to 
take for the nucleon mass the average value, so that we are left only 
with Eq.~(\ref{muqa}) and hence, $\nu_u=\nu_d$. As a starting point, this 
approximation would go along the Walecka model which, with the most common 
parametrizations, does not distinguish the masses of protons and neutrons.
The situation would be different for problems where the neutron-proton
mass difference is the main issue, such as those discussed in 
Ref.~\cite{STMnMp}. We stress that for a fixed $T$ and $\sigma$ the 
quark distribution functions only depend on the  effective chemical 
potentials $\nu_q$, which are determined by the constraints in Eqs.~(\ref{muqa})
and (\ref{muqa1}). Therefore, just as it happens at zero temperature, 
also at finite temperature the effective mass $M^*_N$ does not depend on 
$\omega$ and $b_{03}$.

Before we proceed, we note that we are not considering density/temperature 
dependence in the zero point energy parametrized by $Z$ because this would 
introduce extra, presently unknown parameters. Blunden and Miller (see first
reference in~\cite{recent}) have proposed changing $Z$ through a linear 
dependence with the scalar field and concluded that for reasonable 
parameter ranges this has little effect.

The total energy density at finite temperature $T$ and at finite 
baryon density $\rho_B$ is
\begin{eqnarray}
{\cal E} &=&\frac{2}{(2\pi)^3}\sum_{i=p,n}\int d^3 k \, \left[\epsilon^*\,
(f_i+\bar f_i)+
{\cal V}_{0i}(f_i-\bar f_i)\right]\nonumber\\
&+& \frac{1}{2}{m_\sigma^2}\sigma^2-\frac{1}{2}{m_\omega^2}
\omega^2-\frac{1}{2}{m_\rho^2} b_{03}^2,
\label{ener1}
\end{eqnarray}
where $f_i$ and $\bar f_i$ are the thermal distribution functions
for the baryons and antibaryons,
\begin{equation}
f_i=\frac{1}{e^{(\epsilon^*-\nu_i)/T}+1} ~~{\rm and}~~
\bar f_i=\frac{1}{e^{(\epsilon^*+\nu_i)/T}+1},
\end{equation}
with $i=p,n$ and $\epsilon^*=(\vec k^2+{M^*_N}^2)^{1/2}$ the effective
nucleon energy,  $\nu_i=\mu_i-{\cal V}_{0i}$ the effective
baryon chemical potential  and ${\cal V}_{0i}=g_\omega \omega+ m_\tau\, 
g_\rho b_{03}\,$  ($m_\tau=\pm 1/2$ respectively for protons and 
neutrons).

The thermodynamic grand potential density is defined as
\begin{eqnarray}
\Omega &=& {\cal E} - T {\cal S} - \sum_{i=p,n} \mu_i\rho_i, 
\label{therm}
\end{eqnarray}
with the entropy density ${\cal S} = S/V$ given by
\begin{eqnarray}
{\cal S} &=& -\sum_{i=p,n}\frac{2}{(2\pi)^3}\int d^3 k 
\Big[ f_i\ln f_i+(1-f_i)\ln (1-f_i)\nonumber \\
&+& \bar f_i\ln \bar f_i+(1-\bar f_i)\ln (1-\bar f_i)\Big] .
\label{entropy}
\end{eqnarray}
The proton or neutron density is given by
\begin{equation}
\rho_{i}=\frac{2}{(2\pi)^3}\int d^3 k ~(f_i-\bar f_i),
\end{equation}
so that the baryon density is $\rho=\rho_p+\rho_n$ and the (third component of) 
isospin density $\rho_3=\rho_p - \rho_n$. The proton fraction is defined as
\begin{equation}
y_p=\frac{\rho_p}{\rho}.
\end{equation}

The pressure is the negative of $\Omega$, which after an integration by parts
can be written as
\begin{eqnarray}
P&=& \frac{1}{3}\sum_{i=p,n}\frac{2}{(2\pi)^3}\int d^3 k 
\frac{{\bf k}^2}{\epsilon^*(k)} (f_i+\bar f_i) \nonumber \\
&-& \frac{1}{2}{m_\sigma^2}\sigma^2 + \frac{1}{2}{m_\omega^2}
\omega^2+\frac{1}{2}{m_\rho^2} b_{03}^2.
\end{eqnarray}
From the above expression the pressure depends explicitly on the 
meson mean fields $\sigma$, $\omega$ and $b_{03}$. It also depends 
on the nucleon effective mass $M^{*}_{N}$
which in turn also depends on  the sigma field (see Eqs.~(\ref{Etot}-\ref{MN})).

At a given temperature and for given $\rho_p$, $\rho_n$, the effective nucleon 
mass is known for given values of the meson fields, once the bag radius $R$ 
and the effective quark chemical potentials $\nu_q$  are calculated by using 
Eqs.~(\ref{MNR}) and (\ref{muqa}) respectively. Maximizing the pressure with 
respect to the fields for a given temperature $T$ and given chemical
potentials $\mu_i$, we obtain the following equations for the $\omega$ and 
$b_{03}$ fields
\begin{equation}
\omega=\frac{g_\omega}{m_\omega^2}\rho,\hspace{1.0cm}
b_{03}=\frac{g_\rho}{2m_\rho^2}\rho_3.
\label{vecflds}
\end{equation}
The $\sigma$ meson field  is determined through
\begin{eqnarray}
\frac{\partial P}{\partial \sigma}=
\left( \frac{\partial P}{\partial M^{*}_{N}} \right)_{\mu_i,T}
\frac{\partial M^{*}_{N}}{\partial \sigma}
+\left(\frac{\partial P}{\partial \sigma}\right)_{M^{*}_{N}}=0.
\label{preseg}
\end{eqnarray}
We next calculate the binodal sections within this model.

\subsection{Binodal section}
Nuclear matter is not stable at all temperatures, proton fractions and 
densities. If the Gibbs energy of a two component phase is lower than the 
Gibbs energy of a one component phase, the system will separate into two 
phases~\cite{barranco,muller}. The stability criteria may be expressed  by the
following relations 
\begin{equation}
C_v=\left(\frac{\partial u}{\partial T}\right)_{\rho, y_p}>0, 
\quad K=9 \left(\frac{\partial P}{\partial \rho}\right)_{T, y_p}>0,
\end{equation}
\begin{equation}
\left(\frac{\partial\mu_p}{\partial y_p}\right)_{T,P}\geq 0
\quad \mbox{and}
\left(\frac{\partial\mu_n}{\partial y_p}\right)_{T,P}\leq 0,
\label{diffu}
\end{equation}
where $u=E/B$ is the energy per particle. These conditions
guarantee thermodynamical stability (positive specific heat $C_v$), 
mechanical stability (positive compression modulus $K$) and diffusive 
stability (in a stable system, energy is required to change the proton 
concentration while the pressure and the temperature are kept constant). 
If any of these three criteria is violated there will occur a phase separation.

The surface of two phase coexistence, the binodal surface, is determined 
imposing the Gibbs conditions, namely, for a given temperature, the pressure 
and chemical potentials are equal for both phases. A discussion of the 
properties of phase separation in different models in terms of
temperature, pressure  and proton fraction is most conveniently  done by 
comparing the respective binodal surfaces. 
Understanding the properties of the mixed phase of nuclear matter  is 
important for  different systems
both in astrophysics and nuclear physics. One example is
the crust composition  of a neutron
star, which near the surface is thought to form a Coulomb lattice of nuclei 
immersed in an electron gas and to be responsible for the glitch 
phenomenon~\cite{pet93,glenn00}. Also, in nuclear physics, highly excited nuclei 
created in  heavy-ion collisions gives rise to multifragmentation which is 
interpreted as a  liquid-gas phase transition~\cite{multif}.

In order to obtain the proton and neutron chemical potentials in the two 
coexisting phases for a fixed pressure, we have used the geometrical 
construction \cite{barranco,muller} with the neutron and proton chemical 
potential isobars as a function of proton fraction. This takes into account 
the diffusive stability conditions Eq.~(\ref{diffu}) and  the Gibbs conditions 
for phase separation.
For a given temperature, the binodal section, which contains points under
the same pressure for different proton fractions, is obtained from the above 
conditions, simultaneously with the following equations:
\begin{equation}
P(\nu_p,\nu_n,M^*)
=P(\nu_p^\prime,\nu_n^\prime,{M^*}^\prime),
\label{pr}
\end{equation}
\begin{equation}
\mu_i(\nu_p,\nu_n,M^*)=\mu_i(\nu_p^\prime,\nu_n^\prime,{M^*}^\prime),~~i=p,n,
\end{equation}
\begin{equation}
\left.\frac{\partial\, P}{\partial \sigma}\right|_\sigma=0, 
~~\quad {\rm and}~~
\left.\frac{\partial\, P}{\partial \sigma}\right|_{\sigma'}=0.
\label{sigma}
\end{equation}
At this point, it is worth mentioning that the construction of the binodal 
section for the QMC is slightly more complicated than for the non-linear 
Walecka model. This is because  Eq. (\ref{MNR}) has to be enforced in the 
numerical calculation for every $\sigma$.

The binodal section is formed by two branches, one corresponding to a gas 
phase and small proton fraction and the other to a liquid phase and large 
proton fraction. It is, therefore,  energetically favorable
for nuclear matter to separate into a liquid phase with a large proton 
fraction (less asymmetric) and a gas phase with a small proton fraction 
(more asymmetric).  This behavior is common to other
relativistic mean-field models and reflects the fact that the contribution 
from the $\rho$-meson gives a term in the energy per particle of the form 
$e_{sym}(\rho_B) \delta^2$ where the asymmetry parameter is
$\delta=(\rho_n-\rho_p)/\rho_B$ and the coefficient $e_{sym}$ increases 
with $\rho_B$.

In the next section we will display the binodal section results and compare 
the same with different  non-linear relativistic models.

\section{Results}
We start by fixing the free-space bag properties. We have used zero quark 
masses only and $R_0 = 0.6$~fm for the bag radius. There are
two unknowns, $Z$ and the bag constant, $B$. These are obtained as usual by
fitting the nucleon mass, $M=939$~MeV and enforcing the stability condition
for the bag. The values obtained for $Z$ and $B$ are displayed in Table~1.
The quark-meson coupling constants $g_\sigma^q$, $g_\omega = 3g_\omega^q$ and
$g_\rho = g_\rho^q$ are fitted to obtain the correct saturation properties of 
nuclear matter, $E_B \equiv   \epsilon/\rho - M = -15.7$~MeV at 
$\rho~=\rho_0=~0.15$~fm$^{-3}$ and $a_{sym}=32.5$ MeV. 
We take the standard values for the meson masses, also shown in Table 1.

In the sequel we will frequently  compare QMC to two different parametrizations
of the non-linear Walecka model (NLWM), namely the NL-3 parametrization with 
non-linear contributions from the scalar meson only \cite{nl3} and the TM-1 
parametrization which includes  quartic terms on scalar and vector meson
\cite{tm1}. 

\begin{figure}
\begin{center}
\includegraphics[width=\linewidth]{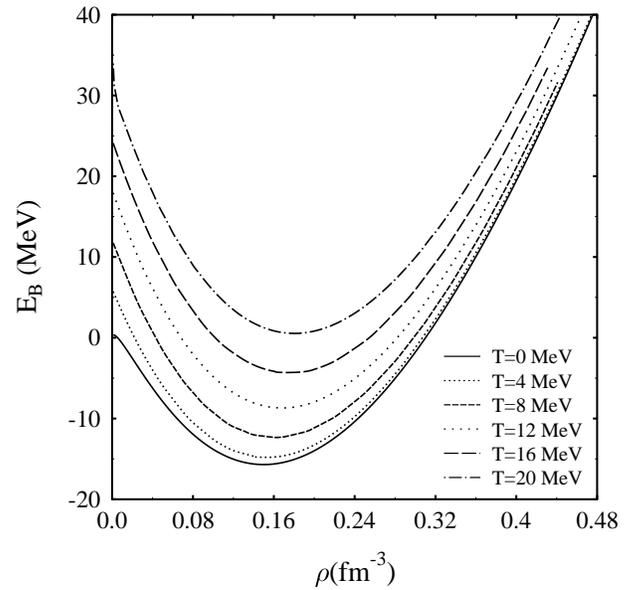}
\end{center}
\caption{Binding energy per nucleon $E_B$ as a function of nuclear
matter density, $\rho$, at temperatures $T$ = 0, 4, 8, 12, 16  
and 20 MeV.}
\label{fig1}
\end{figure}

\begin{figure}
\begin{center}
\includegraphics[width=\linewidth]{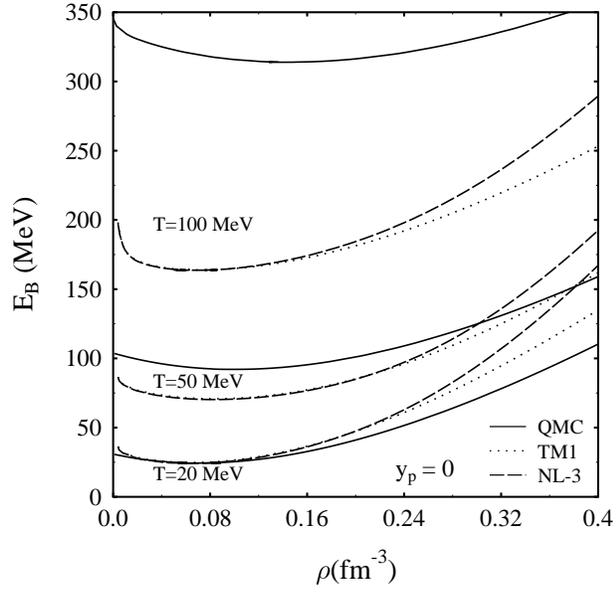}
\end{center}
\caption{Binding energy per nucleon $E_B$ as a function of nuclear
matter density, $\rho$, for neutron matter at temperatures $T$ = 20, 50 and
100 MeV respectively.}
\label{fig1a}
\end{figure}

We first plot the energy per baryon as a function of the baryon 
density in Fig.~\ref{fig1} for symmetric nuclear matter, $y_p=0.5$, 
for temperatures 0, 4, 8, 12, 16  and 20 MeV.
As expected, this functional at zero temperature has a minimum
at the nuclear matter saturation density, $\rho_0$, corresponding
to a binding energy of $-15.7$ MeV. As the temperature increases the
minimum shifts towards higher densities. This may be understood from the
fact that to compensate for the larger kinetic energy a larger value of
$\rho$ is needed to give the minimum.  For larger densities the nuclear 
repulsion effects take
over and  again energy increases. For higher temperatures, the minimum 
of the curves become positive, as in the usual NLWM. In Fig.~\ref{fig1a},
we plot the binding energy for neutron matter for temperatures  20, 50 and
100 MeV and compare the results with TM-1 and NL-3 parametrizations of the NLWM.
At higher temperatures, the QMC model gives a larger binding energy as
compared to the Walecka models. The reason for this is  the contribution
for the thermal energy of the nucleon bag which is absent in the Walecka model.

\begin{figure}
\begin{center}
\includegraphics[width=\linewidth]{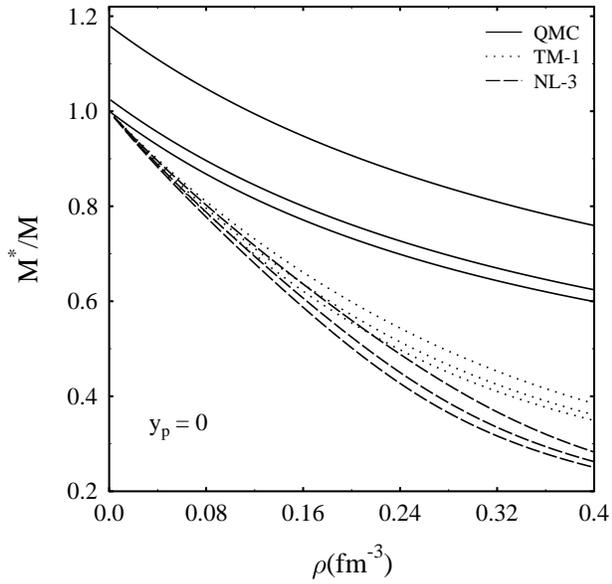}
\end{center}
\caption{Effective mass of the nucleon versus density at temperatures 20, 50 
and 100 MeV in QMC, NL-3 and TM-1 parameters.}
\label{fig4}
\end{figure}

This can be clearly seen from Fig.~\ref{fig4}, where the baryon effective masses 
are plotted as a function of   density for different 
temperatures. In the same figure, we have also shown the results of the
parameter set NL-3 and TM-1  for comparison.
The value of $M^*$  increases with temperature. We have checked that 
at zero density and  temperature  $\sim$ 40 MeV, the QMC effective mass
increases from 939 MeV in contrast to the calculation of hot nuclear matter
using Walecka model. The reason for this is that the sigma field is not strong at
higher temperatures. Also, there is a significant contribution to the 
effective mass arising from the thermal excitation of the quarks inside 
the nucleon bag, which  adds to the mass of the nucleon at higher temperatures.
This contribution which is  absent in the Walecka model calculation, appears 
dominant over the contribution from the sigma field and the net effect is a 
rise of the effective mass. The behavior of the effective mass 
at high temperature 
obtained here is contrary to the results presented in Ref.~\cite{song}, where 
temperature was introduced only at the hadron level, and therefore the behavior 
of the effective mass with temperature not-surprisingly is equivalent to the 
results of Walecka-type of models.

\begin{figure}
\begin{center}
\includegraphics[width=\linewidth]{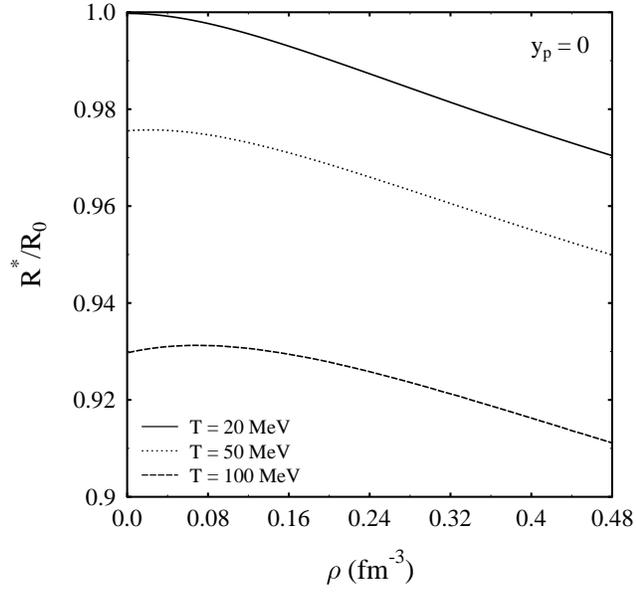}
\end{center}
\caption{The effective bag radius as a function of nuclear matter 
density $\rho$ for temperatures $T$ = 20, 50 and 100 MeV respectively.}
\label{newradius}
\end{figure}

In \ref{newradius} we plot the effective radius in units of the free nucleon
radius $R_0=0.6 $ fm as a function of the baryon density for
temperatures 20, 50 and 100 MeV. It is obesrved that the nucleon bag
shrinks with the increase of temperature. A similar behaviour is also 
predicted with the MQMC model \cite{temp}. Such an effect is
also encountered using the thermal skyrmion \cite{skyrmion} where a
baryon shrinks with temperature.  
We have also calculated the  effective mass  of the nucleon as a 
function of density at different temperatures for the QMC and the MQMC models.
As already discussed in \cite{jin96,mj98}  the effective mass in 
the MQMC decreases faster with $\rho$ than in
the QMC model. However, also in the MQMC there is and increase of the 
effective mass with temperature.
A comparison between the QMC model and different versions of the 
modified MQMC has been performed in Ref. \cite{mj98}. One important 
conclusion was that, contrary to the QMC
result, the bag radius increases
with density for all MQMC models. For densities not much larger than 
nuclear matter
saturation density $\rho_0$ the bags start to overlap which implies a 
breakdown of the
model. On the other hand the MQMC models contain sufficient free parameters 
which allow a good
reproduction of the ground state properties of nuclei. In the present work 
we were interested in
describing propeties of nuclear matter for
densities which go  beyond the saturation density and therefore have chosen to
consider the QMC model.

\begin{figure}
\begin{center}
\includegraphics[width=\linewidth]{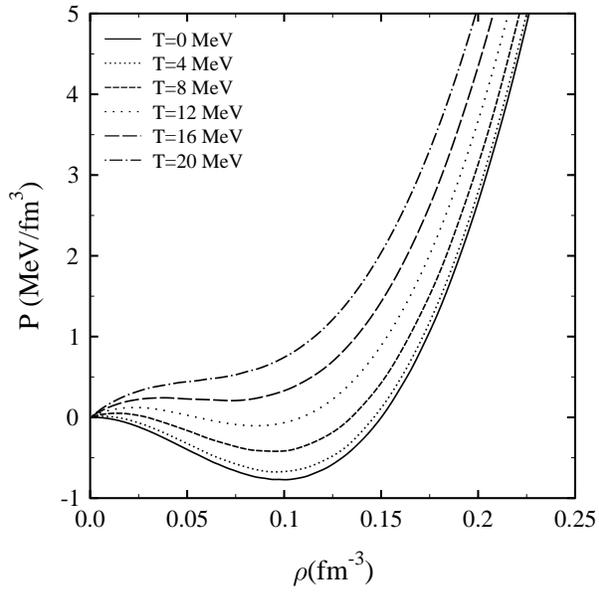}
\end{center}
\caption{Pressure P as a function of nuclear matter density, $\rho$, for
symmetric matter at
temperatures $T$ = 0, 4, 8, 12, 16 and 20 MeV respectively.}
\label{fig2}
\end{figure}

The possible existence of a liquid-gas phase transition is determined 
by the pressure. We plot this quantity as a function of the baryon density, 
$\rho$, for low temperatures in Fig.~\ref{fig2} for symmetric nuclear matter.
At zero temperature, the pressure decreases with density, reaches
a minimum, then increases and passes through $P=0$ at $\rho=\rho_0$,
where the binding energy per nucleon is a minimum. A decrease of the 
pressure with density corresponds to a negative compressibility, 
$K=9 (\partial P/\partial \rho)$,  and is a sign of mechanical instability.
As the temperature increases the region of mechanical instability decreases.
At T=17.7 MeV, the pocket 
disappears. This corresponds to the critical temperature defined by 
$(\partial P/\partial\rho)_{T,y_p} =0=(\partial^2 P/\partial \rho^2)_{T,y_p}=0$,
above which the liquid-gas phase transition is continuous. It is comparable to 
the values for  the critical temperature obtained with Skyrme 
interactions~\cite{jmz84} or relativistic models~\cite{muller}.

\begin{figure}
\begin{center}
\includegraphics[width=\linewidth]{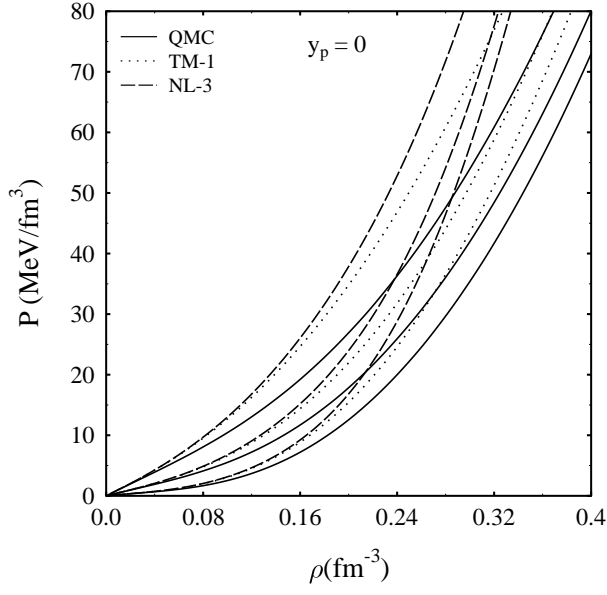}
\end{center}
\caption{Pressure P as a function of nuclear matter density, $\rho$, for
neutron matter at temperatures $T$ = 20, 50 and 100 MeV, respectively, 
from bottom to top. Note that the QMC model gives softer equation of state.}
\label{fig2a}
\end{figure}

In Fig.~\ref{fig2a}, we plot the pressure as function
of density for temperatures 20, 50 and 100 MeV. In the same figure we 
have also shown the  results obtained with the NLWM of TM-1 and NL-3 parameter 
sets to compare. It is found that the QMC model gives a softer EOS,
which is true also for any proton fraction. Although QMC gives a higher energy 
per particle than NL-3 and TM-1, the pressure, which corresponds
to the derivative of the energy per particle with respect to density, does not 
increase so fast with $\rho$.

\begin{figure}
\begin{center}
\includegraphics[width=\linewidth]{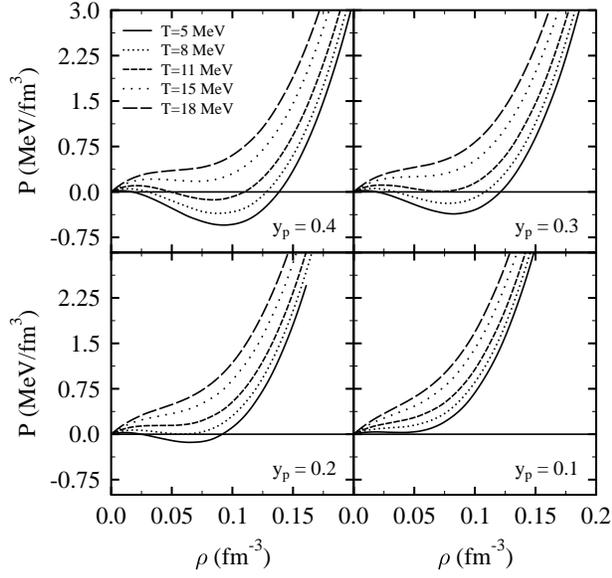}
\end{center}
\caption{Pressure P as a function of nuclear matter density, $\rho$, at
different temperatures and different proton fractions.}
\label{fig2b}
\end{figure}

\begin{figure}
\begin{center}
\includegraphics[width=\linewidth]{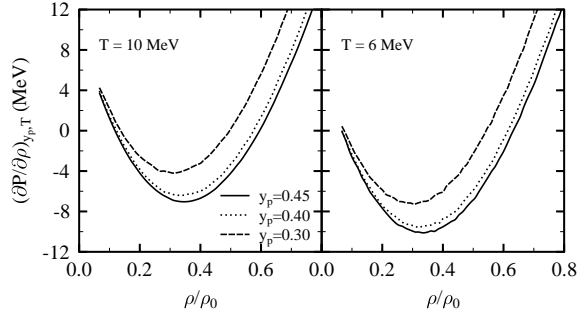}
\end{center}
\caption{$\partial P/\partial \rho$ as a function of the proton fraction.}
\label{dpdr}
\end{figure}

In Fig.~\ref{fig2b} the  pressure is given  as a function of the baryon density, 
$\rho$, for different temperatures and different proton fractions.  The region 
of mechanical instability decreases both with the increase of temperature
and the decrease of the proton fraction. This is clearly seen
in Fig.~\ref{dpdr}, 
where the quantity $(\partial P/\partial \rho)_{T,y_p}$ is shown for two 
different temperatures and three different proton fractions. The curve 
corresponding to $T=10$ MeV and $y_p=0.3$ has  the smallest region with a 
negative value of the derivative $(\partial P/\partial \rho)_{T,y_p}$. 

\begin{figure}
\begin{center}
\includegraphics[width=\linewidth]{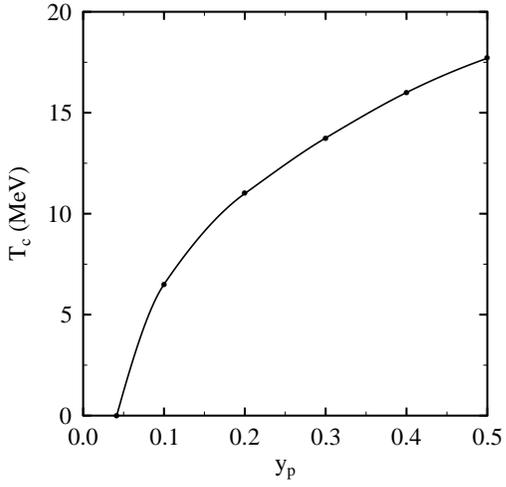}
\end{center}
\caption{Critical temperature as a function of the proton fraction.}
\label{fig2c}
\end{figure}

The behavior of the critical temperature with the proton fraction is shown in
Fig.~\ref{fig2c}. It decreases rapidly with $y_p$ and is  going to zero for  
$y_p=0.041$. Although there will be no mechanical instabilities above $T_c$, 
chemical instabilities can still develop above $T_c$.  The present  results 
are comparable to the ones discussed in Refs.~\cite{muller,kuo,li}. 

\begin{figure}
\begin{center}
\includegraphics[width=\linewidth]{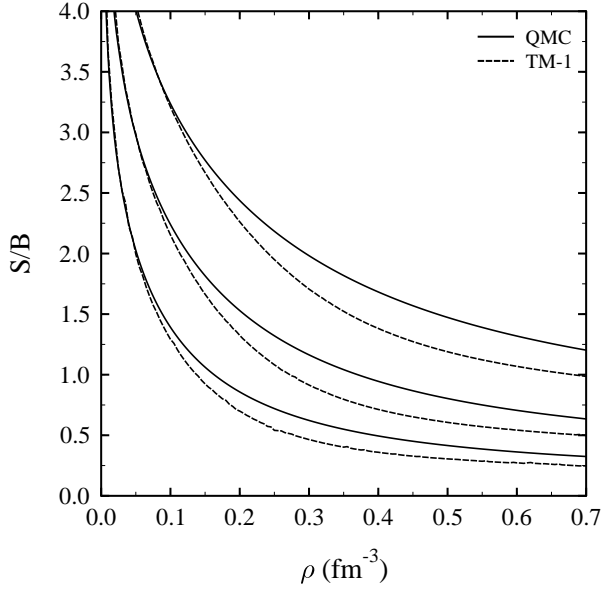}
\end{center}
\caption{Entropy per baryon, $S/B$, calculated within QMC and TM-1, as a 
function of nuclear matter density, $\rho$, for symmetric matter at 
temperatures $T$ = 10, 20 and 40 MeV from bottom to top.}
\label{fig3}
\end{figure}

The entropy per baryon is an important quantity to describe a 
collapsing massive star. During the whole process of the collapse the 
entropy per baryon, including the contribution of baryons and leptons, 
remains low: it is of the  order of $S/\rho\sim1$ initially, increases slightly 
before neutrino trapping and remains constant, $S/\rho\sim1-2$, 
afterwards~\cite{shapiro}.  According to different models the contribution 
from the baryons alone changes from 80-90\% at the  saturation density to 
50-65\% at 10 times that density \cite{prakash}.  In Fig.~\ref{fig3} the 
entropy per baryon is plotted and compared with the corresponding quantity 
calculated within the TM-1 parametrization of NLWM. We conclude that QMC 
gives values compatible with the ones discussed in Ref.~\cite{prakash}. 
The entropy per baryon in QMC  decreases slower than in  TM-1.

In Fig. 11, we plot the scalar and vector potentials  as a function of density
for  T=8 MeV and $y_p=0.4$. In the same figure we have also shown 
the non-linear Walecka model results.  The $\omega$ and $\sigma$ contributions
in the QMC are both weaker than the corresponding contributions in the NL-3 and TM-1
parametrizations of the NLWM. As a consequence, the QMC effective mass changes
less with density than the TM-1  and NL-3 effective masses. This can also be 
confirmed in Fig.~3. A second consequence is a softer EOS due to the weaker 
omega field.

Isobars of nuclear pressure are plotted on the $\mu_p$ and $\mu_n$ space  
in Fig. 12 for $T=6$ MeV for two different pressures, $P=0.1$ and 
$0.063$ MeV/fm$^3$. 
The diffusively unstable regions can be seen clearly
in chemical potential isobars in this figure. According to the  inequality 
(\ref{diffu}), the region of negative (positive) slope for $\mu_p~ (\mu_n)$ 
is unstable. Violation of the stability criteria is an indication of phase 
separation. The surface of the two phase coexistence in the $(p,T,y_p)$ space, 
the binodal surface, is obtained from conditions (\ref{pr})-(\ref{sigma}).
In Fig. 13 we plot the binodal sections for the  QMC and the NL-3 and TM-1 
parametrizations, at T= 6 MeV.  The critical point occurs for similar values 
of the pressure and proton fraction in all three models. For the TM-1 
parametrization,  the critical pressure is 10\% larger than the QMC critical 
pressure. On the other hand  the proton fraction for NL-3 is  $\sim$ 10\% larger 
than the corresponding values for QMC. The largest difference between the models 
is in the shape: the region of configuration space where the phase separation 
is favorable is smaller in the QMC and occurs at smaller values of $y_p$ in 
the liquid phase. As a consequence   chemical instability occurs for smaller 
isospin asymmetries,  which may have consequences in isospin 
fractionation~\cite{li}. The finite temperature results  agree with the ones
obtained within  QMC at zero temperature, \cite{krein00}.  For larger 
temperatures, the  binodal has a similar shape but extends to smaller isospin 
asymmetries~\cite{muller,mp}.  

\section{Summary}
We now  summarize the results and conclusions of the present work. 
We have studied
asymmetric nuclear matter at finite temperature using the QMC model which
incorporates explicitly quark degrees of freedom. The results for the EOS were
compared with the two non-linear Walecka models, namely NL-3 and TM-1 
parametrizations. The mean effective fields $\sigma$, $\omega$ and $\rho$  
are determined from the minimization of the thermodynamical potential,
and the temperature dependent effective bag radius was calculated from the 
minimization of the effective mass of the nucleon mass of the bag. The thermal
contributions of the quarks, which are absent in the  Walecka 
model, are dominant
and lead to a rise of the effective nucleon mass at finite temperature. 
This is contrary to the results presented in Ref.~\cite{song}, where
temperature was introduced only at the hadron level, and therefore the behavior 
of the effective mass with temperature is equivalent to the results of  
Walecka-type models~\cite{qhd}. In the present calculation, the effective 
radius of the nucleon bag shrinks with increase of the temperature. 

The equation of state as derived here is softer than the ones obtained within 
the non-linear Walecka model for the NL-3 and TM-1 parametrizations, for all 
temperatures. The region of mechanical instability decreases with the increase 
of $T$ and decrease of $y_p$. Also, the critical temperature decreases with  
the proton fraction, from a maximum value of $T=17.7$ MeV at $y_p=0.5$ 
to $T=0$ MeV at $y_p=0.041$.  We have  shown that the potentials in the 
QMC are weaker than the corresponding ones in the NLWM for the NL-3 and TM-1 
parametrizations, which imply a softer EOS and a weaker change of the
effective mass with $\rho$.

We have also constructed and studied the binodal sections which is essential 
for studying  the liquid-gas phase transition and to understand under which 
conditions chemical instabilities develop, leading to isospin fractionation. 
It is clear from the binodal that the system prefers to separate into regions 
of higher density and smaller isospin asymmetry and regions of lower  
density and large isospin asymmetry~\cite{muller,li,mp}.  The phase separation 
in the QMC occurs at smaller isospin asymmetry, corresponding to lower values 
of $y_p$, than the  ones predicted by the NL-3 and TM-1 parametrizations of 
the NLWM. 

Extensions of the formalism presented here to droplet formation are currently 
under process.

\acknowledgments{Work partially supported by CAPES, CNPq and  FAPESP (Brazil), 
and FEDER, GRICES  and CFT (Portugal) under the contract POCTI/35308/FIS/2000.
The facilities offered by the Center for Computational Physics, University 
of Coimbra are warmly acknowledged.}

\newpage
\begin{table}
\begin{ruledtabular} 
\caption{Parameters used in the calculation for $R_0=0.6$ fm. All masses and bag
pressure are in MeV.}
\begin{tabular}{cccccccccc}
$m_q$& $B^{1/4}$ & $Z$ & $m_\sigma$
& $m_\omega$& $m_\rho$&
$g_\omega$&$g_\sigma^q$&$g_\rho$\\ \hline
0 & 211.3 & 3.987 & 550 & 783& 763 &8.9539&5.981&8.655 \\
\end{tabular}
\end{ruledtabular} 
\end{table}
\newpage
\begin{figure}
\begin{center}
\includegraphics[width=\linewidth]{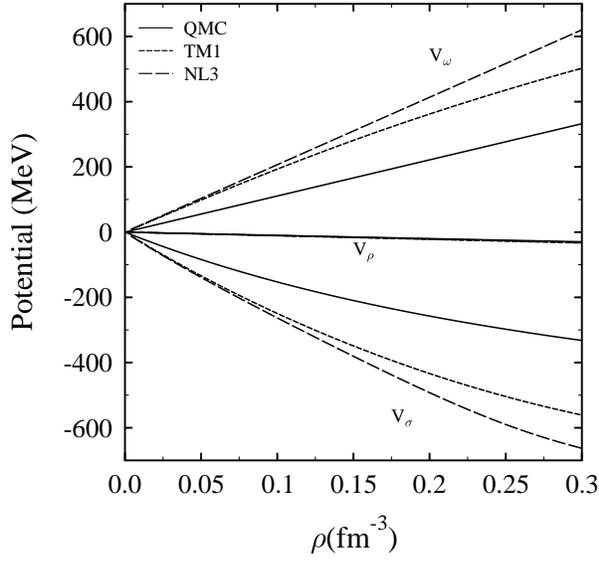}
\end{center}
\caption{Potentials versus nuclear matter density, $\rho$, at temperature
= 8 MeV and $y_p=0.4$ in QMC, NL-3 and TM-1 parameters.}
\label{fig6}
\end{figure}
\begin{figure}
\begin{center}
\includegraphics[width=\linewidth]{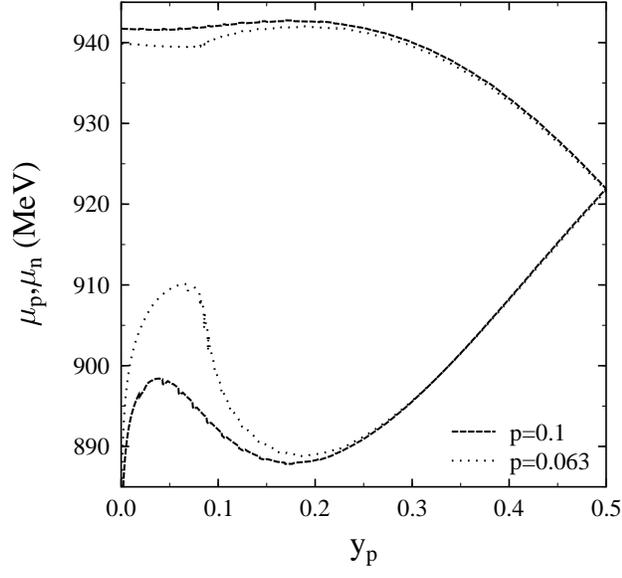}
\end{center}
\caption{Nuclear pressure isobars for the proton (lower) and neutron (upper) 
chemical potentials as a function of proton fraction for $P=0.1$ and 
$0.063$ MeV/fm$^3$. }
\label{fig7}
\end{figure}
\begin{figure}
\begin{center}
\includegraphics[width=\linewidth]{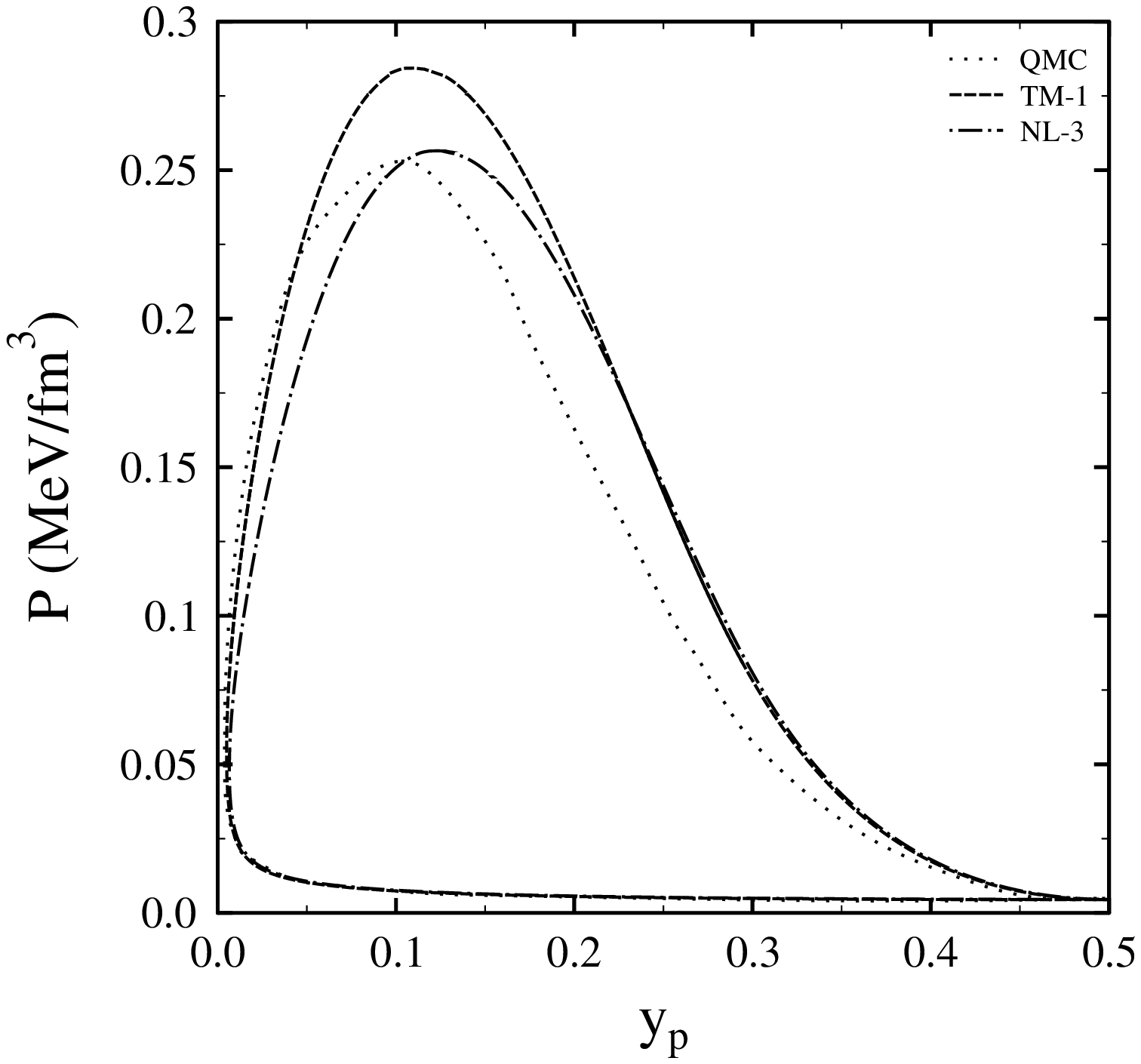}
\end{center}
\caption{The binodal section at T=6 MeV for QMC, NL-3 and TM-1 }
\label{fig12}
\end{figure}

\begin{thebibliography}{99}
%
\bibitem{prakash} M. Prakash, I. Bombaci, M. Prakash, P. J. Ellis, J. M. Lattimer 
and R. Knorren, Phys. Rep. {\bf 280}, 1 (1997).
%
\bibitem{NA50} M. C. Abreu {\em et al}, Phys. Lett. {\bf B 477}, 28 (2000).
%
\bibitem{multif} J. A. Hauger {\em et al}, Phys. Rev. Lett. {\bf 77}, 235 (1996); 
J. Phochodzalla  {\em et al}, Phys. Rev. Lett. {\bf 75}, 1040 (1995).
%
\bibitem{pethick}C.J. Pethick, D.G. Ravenhall, and C.P. Lorenz, Nucl. Phys.
{\bf A584}, 675 (1995).
%
\bibitem{pet93}C. P. Lorenz, D. G. Ravenhall, and C. J. Pethick Phys. Rev. Lett. 
{\bf 70}, 379 (1993). 
%
\bibitem{barranco}M. Barranco and  J.R. Buchler, Phys. Rev. C {\bf 22}, 
1729 (1980).
%
\bibitem{muller}H. M\"uller and  B.D. Serot, Phys. Rev. C {\bf 52}, 
2072 (1995).
%
\bibitem{mp}D.P. Menezes and  C. Providencia, Nucl. Phys. {\bf A650}, 
283 (1999); Phys. Rev. C {\bf 60}, 024313 (1999).
%
\bibitem{guichon}P. A. M. Guichon, Phys. Lett. {\bf B200}, 235 (1988).
%
\bibitem{ST}K. Saito and A.W. Thomas, Phys. Lett. {\bf B327}, 9 (1994);
{\bf B335}, 17 (1994); {\bf B363}, 157 (1995); Phys. Rev. C {\bf 52}, 2789
(1995); P.A.M. Guichon, K. Saito, E. Rodionov, and A.W. Thomas, Nucl. Phys.
{\bf A601} 349 (1996);  K. Saito, K. Tsushima, and A.W. Thomas, Nucl. Phys.
{\bf A609}, 339 (1996); Phys. Rev. C {\bf 55}, 2637 (1997); Phys. Lett. 
{\bf B406}, 287 (1997).
%
\bibitem{qhd} B.D. Serot and J.D. Walecka, Int. J. Mod. Phys. E {\bf 8}, 
515 (1997); J.D. Walecka, Ann. of Phys. (N.Y.) {\bf 83}, 491 (1974).
%
%
\bibitem{ffactors}G. Krein, A.W. Thomas, and K. Tsushima,
Nucl. Phys. {\bf A650}, 313 (1999); M.E. Bracco, G. Krein, and
M. Nielsen, Phys. Lett. {\bf B432} 258 (1998).
%

\bibitem{recent}P. G. Blunden and G.A. Miller, Phys. Rev. C
{\bf 54}, 359 (1996); N. Barnea and T.S. Walhout, Nucl. Phys. {\bf A677}, 
367 (2000); H. Shen and H. Toki, Phys. Rev. C {\bf 61}, 045205 (2000);
P.K. Panda, R. Sahu, C. Das, Phys. Rev. C {\bf 60}, 38801 (1999);
P.K. Panda, M.E. Bracco, M. Chiapparini, E. Conte, and G. Krein,
Phys. Rev. C {\bf 65}, 065206 (2002); P.K. Panda, and F.L. Braghin, Phys. Rev.
C {\bf 66}, 055207 (2002).
%
\bibitem{temp}P.K. Panda, A. Mishra, J.M. Eisenberg, W. Greiner,
Phys. Rev. C {\bf 56}, 3134 (1997); I. Zakout, H.R. Jaqaman,
Phys. Rev. C {\bf 59}, 962 (1999).
%
\bibitem{song} H. Q. Song and R. K. Su, Phys. Lett. {\bf B358}, 179 (1995).
%
\bibitem{krein00}G. Krein, D.P. Menezes, M. Nielsen,
and  C. Providencia, Nucl. Phys. {\bf A674}, 125 (2000).
%

\bibitem{jin96}X. Jin and B.K. Jennings, Phys. Lett. {\bf B374}, 13
(1996); Phys. Rev. C {\bf 54}, 1427 (1996).
%
\bibitem{mj98} H.  Mueller and  B. K. Jennings, Nucl.Phys. A {\bf 640} 
(1998) 55-76; Nucl. Phys. A  {\bf
    626}, 966 (1997).


\bibitem{EFT} R. J. Furnstahl and B.D. Serot, Comm. Nucl. Part. Phys. {\bf 2}
A23 (2000); R.J. Furnstahl, B.D. Serot and H.-B. Tang, Nucl. Phys. {\bf A598}, 
539 (1996); Nucl. Phys. {\bf A614}, 441 (1997); Nucl. Phys. {\bf A618}, 
446 (1997).
% 
\bibitem{mahaux87} C. H. Johnson, D. J. Horen and C. Mahaux, Phs. Rev. C 
{\bf 36}, 2252 (1987).
%
\bibitem{nl3} G. A. Lalazissis, J. K\"onig and P. Ring,
Phys. Rev. C {\bf 55}, 540 (1997).
%
\bibitem{tm1} K. Sumiyoshi, H. Kuwabara, H. Toki, Nucl. Phys. {\bf A581}, 725
(1995).
%
\bibitem{STMnMp}K. Saito and A. W. Thomas, Phys. Lett. {\bf B335}, 17 (1994);
T. Hatsuda, E.M. Henley, Th. Meissner and G. Krein, Phys. Rev. C {\bf 49},
452 (1994); E.M. Henley and G. Krein, Phys. Rev. Lett. {\bf 62}, 2586 (1989).
%
\bibitem{glenn00} N. K. Glendenning, Compact Stars, Springer-Verlag, New York, 
2000.
%
\bibitem{skyrmion}J. Dey and J.M. Eisenberg, Phys. Lett. {\bf B 334}, 290 (1994);
M.A. Nowak and I. Zahed, Phys. Lett. {\bf B 230}, 108 (1989); K.J. Eskola and
K. Kajantie, Z. Phys. {\bf C 44} 347 (1989).
%
\bibitem{jmz84} H. R. Jaqaman, A. Z. Mekjian, and L. Zamick, Phys. Rev. C 
{\bf 29}, 2067 (1984).
%
\bibitem{kuo} S. Ray, J. Shamanna, T. T. S. Kuo, Phys.Lett. {\bf B392}, 7 (1997).
%
\bibitem{li} Bao-An Li, C.M. Ko, and W. Bauer, Int. J. Mod. Phys. E
{\bf 7} 147 (1998); B.-A. Li, A. T. Sustich, Matt Tilley, B. Zhang, Nucl. Phys. 
{\bf A699}, 493 (2002).
%
\bibitem{shapiro} S. L. Shapiro and S. A.  Teukolsky, {\it Black holes, white 
dwarfs and neutron stars, the physics of compact objects}, Wiley, 
New York, 1983.
\end{thebibliography}
\end{document}